\documentclass[a4paper,11pt]{article}
\begin{document}
\def\baselinestretch{1.5}
{\def\title{THERMODYNAMICS WITHIN THE FRAMEWORK OF CLASSICAL
MECHANICS}
\def\author{V.M. Somsikov}
\def\correspondence{Prof. V.M. Somsikov}
\def\correspondenceshort{Prof. Somsikov}
\def\date{\today}
\def\addra{Laboratory of Physics of the geogeliocosmic relation}
\def\addrb{Institute of Ionosphere}
\def\addrc{Kamenskoe Plato}
\def\addrd{Almaty, 480020, Kazakhstan}
\def\tel{~~  ~~~+8-3272-548074~~~~~~~~~~~~~~~~~~~~~}
\def\fax{~~  ~~~+8-3272-658085~~~~~~~~~~~~~~~~~~~~~}
\def\email{~~  ~~ nes@kaznet.kz ~~~~~~~~~~}}
\small
\title {THERMODYNAMICS WITHIN THE FRAMEWORK OF CLASSICAL MECHANICS}
\author{Vyacheslav Somsikov }
\date{\it{Laboratory of Physics of the geogeliocosmic relation,\\
Institute of Ionosphere, Almaty, 480020, Kazakhstan}} \maketitle
\begin{abstract}

The approach to a substantiation of thermodynamics is offered. A
conservative system of interacting elements, which is not in
equilibrium, is used as a model. This system is then split into
small subsystems that are accepted as being in equilibrium. Based
on the D'Alambert equation for a subsystem the generalized
Liouville equation is obtained. A necessary condition for
irreversibility is determined. This condition is dependence the
forces of interaction of subsystems on relative velocities. The
equation of motion of subsystems of potentially interacting
elements is obtained. The non-potentiality of the forces of
interaction of the subsystems consisting of potentially
interacting elements is proved. The mechanism of occurrence of
irreversible dynamics is offered. The formula that expresses the
entropy through the forces of interaction of subsystems is
obtained. The theoretical link between classical mechanics and
thermodynamics is analyzed.
\\

\end{abstract}
\bigskip

\date{\it{1. Introduction.}} \maketitle

The theoretical link between classical mechanics and
thermodynamics can be constructed by investigation of the
irreversibility mechanism. Irreversibility is an essential part of
the second law of thermodynamics. According to this law there is
function $'S'$ named entropy, which can only grow for isolated
systems, achieving a maximum at a state of equilibrium. But this
is in contradiction with reversibility of the Newton equation and
potentiality of the fundamental forces [Hooft, 1980]. The great
importance of the irreversibility problem for fundamental physics
explains its big popularity among the physicists. The history of
its solution was very extensive and sometimes, dramatic.
Therefore, let us reference only few those works, which precisely
enough gives a clear picture of the state in the problem of
irreversibility [Cohen, 1998; Lebowitz, 1993; Petrosky {\&}
Prigogine, 1997; Zaslavsky, 1999].

A first attempt to resolve this contradiction has been done by
Boltzmann. From the $H$-theorem it follows that many-body systems
should equilibrate. But for obtaining this result, Boltzmann had
used principles of probability. Therefore the contradiction was
not overcome.

For overcoming this problem, it was suggested to try to create an
expanded formalism of open systems within the framework of
classical mechanics laws [Petrosky {\&} Prigogine, 1997]. It
turned out that such formalism appears in the process of solving
the problem of irreversibility for a hard-disks system due to
refusal from conservatism of systems and potentiality of their
interactions [Somsikov {\&} Matesov, 2001; Somsikov 2001, 2004a,
2004b, 2004c].

The substantiation of the mechanism of equilibration for
non-dissipative hard-disks system was based on the dependence of
force of disks interaction on their relative velocities, and on
the necessary condition for irreversibility [Somsikov, 2004a]. The
dependence of the force of disks interaction on the velocities
followed from their equation of motion. The existence of
irreversibility condition followed from the general Liouville
equation. But in natural systems the forces of interacting of the
elementary particles are potential [Hooft, 1980] and therefore the
equation of motion is reversible. Thus there is a key question on
a way of a substantiation of thermodynamics: is it possible to
connect the fact of the potentiality of interaction elements in
the system with irreversibility which in real systems exists? The
search of the answer on this question is the purpose of this work.

The investigation is based on the similar method which for a
hard-disks system was used. A conservative system of interacting
elements, which is not in equilibrium, is prepared. This system is
then split into small subsystems that are accepted as being in
equilibrium. The subsystem dynamics under condition of their
interactions is analyzed on the basis of classical mechanical
laws.

The work was carried out in the following way. First of all, based
on the equation of motion for disks, we will show that the forces
between the selected subsystems, which we called the generalized
forces, are dependent on the velocities. This result has led us to
the key point that on the base of usual canonical Hamilton and
Liouville equations, the fundamental problem of irreversibility
cannot be solved. Instead we have used so-called generalized
Hamilton and Liouville equations [Somsikov {\&} Matesov, 2001].
Based on this, we discover the condition of irreversible dynamics.
This condition follows from the generalized Liouville equation.

Then the dynamics of the systems constructed from potentially
interaction elements is analyzed. We will call these systems as
Newtonian systems. The equation of motion of the subsystem is
obtained. Based on this equation we answer the question of why and
how the velocity dependence of generalized forces between
subsystems appears when the forces between the elements are
potential.

Then based on the equation of motion of the subsystem the
essential link between thermodynamics and classical mechanics is
analyzed.

A formula, which expresses the entropy through the work of
generalized forces, is obtained. This formula is determined by the
fact that the energy of subsystem interaction is transformed into
internal energy as a result of the work done by the generalized
forces in an irreversible way.

\date{\it{2. Irreversibility in a hard-disks system.}} \maketitle

The study of a hard-disk system is based on the equation of motion
for disks. This equation is deduced with help of the matrix of
pair collisions. In the complex plane this matrix is given
[Somsikov, 2001]:

$S_{kj}=\left(\begin{array}{cc} a & -i b
\\ -i b & a\end{array} \right)$ (a),

where $a=d_{kj}\exp(i\vartheta_{kj})$; $b=\beta
\exp(i\vartheta_{kj})$; $d_{kj}=cos\vartheta_{kj}$;
$\beta=sin\vartheta_{kj}$; $i$ is the imaginary unit; $k$ and $j$
numbers of colliding disks; $d_{kj}$ is the impact parameter (IP),
determined by the distance between the centers of colliding disks
in a Cartesian plane system of coordinates with axes of $x$ and
$y$, in which the $k$-disk swoops on the lying $j$- disk along the
$x$ - axis. The scattering angle $\vartheta_{kj}$ varies from $0$
to $\pi$. In consequence of collision the transformation of disks
velocities can be presented in such form: $V_{kj}^{+}=S_{kj}
V_{kj}^{-}$ (a), where $V_{kj}^{-}$ and $V_{kj}^{+}$ - are
bivectors of velocities of $k$ and $j$ - disks before $(-)$, and
after $(+)$ collisions, correspondingly; $V_{kj}$=$\{V_k,V_j\}$,
${V}_j=V_{jx}+iV_{jy}$ - are complex velocities of the incident
disk and the disk - target with corresponding components to the
$x$- and $y$- axes. The collisions are considered to be central,
and friction is neglected. Masses and diameters of disks, $"d"$,
are accepted to be equal to $1$. Boundary conditions are given as
either periodical or in form of hard walls. From (a) we can obtain
equations for the change of velocities of colliding disks:
\begin{equation}
{\left(\begin{array}{c} {\delta V_k}\\ {\delta V_j}
\end{array} \right )
=\varphi_{kj} \left( \begin{array}{c} \Delta_{kj}^{-} \\
-\Delta_{kj}^{-}\end{array} \right )}.\label{eqn1}
\end{equation}
Here, $\Delta_{kj}=V_k-V_j$ are relative velocities, $\delta
V_k=V_k^{+}-V_k^{-}$, and $\delta V_j=V_j^{+}-V_j^{-}$ - are
changes of disks velocities in consequence of collisions,
$\varphi_{kj}=i\beta e^{i\vartheta_{kj}}$. \\That is, Eq. (1) can
be presented in the differential form [Somsikov, 2001]:
\begin{equation}
\dot{V}_k=\Phi_{kj}\delta (\psi_{kj}(t))\Delta_{kj}\label{eqn2}
\end{equation}
where $\psi_{kj}=[1-|l_{kj}|]/|\Delta_{kj}|$;
$\delta(\psi_{kj})$-delta function; $l_{kj}(t)=z_{kj}^0+\int
\limits_{0}^{t}\Delta_{kj}{dt}$ are distances between centers of
colliding disks; $z_{kj}^0=z_k^0-z_j^0$, $z_k^0$ and $ z_j^0$ -
are initial values of disks coordinates;
$\Phi_{kj}=i(l_{kj}\Delta_{kj})/(|l_{kj}||\Delta_{kj}|)$.

The Eq. (2) determines a redistribution of kinetic energy between
the colliding disks. It is not a Newtonian equation because the
forces between the colliding disks depend on their relative
velocities. Hence, for the analysis of systems of disks it is
impossible to use the canonical Hamilton equation [Lanczos, 1962].
So, we get the generalized Hamilton equation to be applied for
studying the subsystem dynamics [Somsikov {\&} Matesov, 2001]:

\begin{equation}
{\frac{\partial{H_p}}{\partial{r_k}}=-\dot{p_k}+F_{k}^p}\label{eqn3}
\end{equation}
\begin{equation}
{\frac{\partial{H_p}}{\partial{p_k}}={V_k}}\label{eqn4}
\end{equation}
These are the general Hamilton equations for the selected
$p$-subsystem. The external forces, which acted on $k$ disks
belong to the $p$-subsystem, presented in a right-hand side on Eq.
(3). These forces are not potential.

Using Eqs. (3,4), we can find the Liouville equation for
$p$-subsystem. For this purpose, let us take a generalized current
vector $J_p=(\dot{r_k},{\dot{p_k}})$ of the $p$-subsystem in a
phase space. From Eqs. (3,4), we find [Somsikov {\&} Matesov,
2001; Somsikov, 2004a]]:
\begin{equation}
{\frac{df_p}{dt}=-f_p\sum\limits_{k=1}^T
\frac{\partial}{\partial{p_k}}F_{k}^p} \label{eqn5}
\end{equation}

Eq. (5) is a Liouville equation for the $p$-subsystem. We can
rewrite it in differential form as: \\$\frac{df_p}{dt}+f_p
div_{\vec{p}}{\vec{F}}=0 $, where $\vec{p}=\{p_k\};
\vec{F}=\{F^{p}_k\}, k=1,2...T $.

The Liouville equation has the formal
solution:\\${f_p=const\cdot{\exp{[-\int\limits_{0}^{t}{(\sum\limits_{k=1}^T
\frac{\partial}{\partial{p_k}}F_{k}^p)}{dt}]}}}$.

The Eq. (5) is obtained from the common reasons. Therefore it is
suitable for any interaction forces of subsystems, as dissipative,
as non-dissipative. For a hard-disks system the energy dissipation
does not exist. There is only a redistribution of kinetic energy
between colliding disks. Thus, Eq. (5) is applicable to analyze
any open nonequilibrium systems, because it takes into account the
energy exchange between subsystems. Therefore it can be used also
for the explanation of irreversibility in a frames of the
classical mechanics laws.

The right side of Eq. (5), $f_p\sum\limits_{k=1}^T
\frac{\partial}{\partial{p_k}}F_{k}^p$, is a similar to the
integral of collisions. It can be obtained using the subsystems
motions equations. For a hard disks system it can be found with
the help of Eq. (2).

In the non-equilibrium system the right term of Eq. (5) is not
zero, because generalized forces are dependent on velocities.
Therefore the relative subsystems velocities are distinguish from
zero. This conclusion is in agreement with the fact, that
subsystems velocities in the non-equilibrium systems are non-zero
[Landau {\&} Lifshits, 1976]. This can mean only, that when the
system goes to equilibrium state, the relative velocities of
subsystems go to zero.

Let us consider the important interrelation between descriptions
of dynamics of separate subsystems and dynamics of system as a
whole. As the expression,
${\sum\limits_{p=1}^R{\sum\limits_{k=1}^T F_{k}^p =0}}$, is
carried out, the next equation for the full system Lagrangian,
$L_R$, will have a place:
\begin{equation}
{\frac{d}{dt}\frac{\partial{L_R}}{\partial{V_k}}-
\frac{\partial{L_R}}{\partial{r_k}}=0}\label{eqn6}
\end{equation}
and the appropriate Liouville equation:
${\frac{\partial{f_R}}{\partial{t}}+{V_k}\frac{\partial{f_R}}
{\partial{r_k}}+\dot{p_k}\frac{\partial{f_R}}{\partial{p_k}}=0}$.
The function, $f_R$, corresponds to the full system. The full
system is conservative. Therefore, we have: ${\sum\limits_{p=1}^R
divJ_p=0}$. This expression is equivalent to the next equality:
${\frac{d}{dt}(\sum\limits_{p=1}^{R}\ln{f_p})}=
\frac{d}{dt}(\ln{\prod\limits_{p=1}^{R}f_p})=
{(\prod\limits_{p=1}^{R}f_p)}^{-1}\frac{d}{dt}(\prod\limits_{p=1}^{R}{f_p})=0$.
So, $\prod\limits_{p=1}^R{f_p}=const$. In an equilibrium state we
have $\prod\limits_{p=1}^R{f_p}=f_R$. Because the equality
$\sum\limits_{p=1}^{R}F_p=0$ is fulfilled during all time, we have
that equality, $\prod\limits_{p=1}^R{f_p}=f_R$, is a motion
integral. It is in agreement with Liouville theorem about
conservation of phase space [Lanczos, 1962].

So, only in two cases the Liouville equation for the whole
non-equilibrium system is in agreement with the general Liouville
equation for selected subsystems: if the condition
$\int\limits_{0}^{t}{(\sum\limits_{k=1}^T\frac{\partial}
{\partial{p_k}}F_{k}^p)}dt\rightarrow{const}$ (c) is satisfied
when $t\rightarrow\infty$, or when,
${(\sum\limits_{k=1}^T}\frac{\partial}{\partial{p_k}}F_{k}^p)$, is
a periodic function of time. The first case corresponds to the
irreversible dynamics, and the second case corresponds to
reversible dynamics [Somsikov, 2004a].

Because the generalized forces for a hard-disks system depended
on velocities, the irreversible dynamics is possible.

Dynamics of strongly rarefied systems of potentially interacting
elements is also described by the Eq. (2). Therefore for those
systems, irreversibility is possible as well [Somsikov, 2004a,
2004b].

We see that the dependence of generalized forces on velocities for
subsystems interaction is a necessary condition for the
irreversibility to occur. So, the question about irreversibility
for Newtonian systems is reduced to that about dependence of the
forces between subsystems on the velocity.

For a hard-disks system and for strongly rarefied system of
potentially interacting elements the presence of irreversibility
is predetermined by Eq. (2). In these systems the interaction
forces between the elements are depending on velocities. Therefore
it is clear that the generalized forces will depend on velocities
as well. But the forces between elements for Newtonian systems are
potential. Therefore it is necessary to answer the question: how
does velocity dependence of generalized force between subsystems
appear when forces between the elements are independent on
velocities. The answer on this question is a purpose of the next
part of the paper. For this aim the equation of motion for
Newtonian subsystems will be obtained.

\date{\it{3. The subsystems equation of motion }}\maketitle

Let us to analyze Newtonian systems. We take a system with energy:
$E_N=T_N+U_N=const$, where
$T_N=\frac{1}{2}\sum\limits_{i=1}^N{{v_i}^2}$ is a kinetic energy;
$U_N(r_{ij})$ is potential energy; $r_{ij}=r_i-r_j$ is the
distance between $i$ and $j$ elements; $N$ is the number of
elements. Masses of elements are accepted to $1$.

The Newton equation of motion for elements is:
\begin{equation}
{\dot{v}_i=-\sum\limits_{i=1,j\neq{i}}^N\frac{\partial}
{\partial{r_{ij}}}U}\label{eqn7}
\end{equation}

It is obviously that the irreversibility in Newtonian systems must
be compatible with reversibility of the Newton equation. But
already for system of two bodies the force acting on bodies has
nonlinear character. It depends on their relative velocity. By
transition in system of coordinates of the center of mass this
nonlinearity manages to be excluded. For three bodies such
exception generally becomes impossible. Therefore this task
generally is not solved. From here follows, that it is impossible
to assert a priori about potentiality of interaction of many-body
systems of under condition of potentiality of pair interactions of
elements. We shall show, that non-potentiality the generalized
forces, i.e. forces of interaction of subsystems, follows directly
from their equations of motion.

Therefore, for the description of evolution of nonequilibrium
system, it is necessary to obtain an equation of subsystem motion,
which determines the generalized forces. For this purpose we will
do the following. In some system of coordinates we represent the
total subsystem energy as the sum of the kinetic energy of
subsystem motion as the whole, $T_n^{tr}$, the kinetic energy of
its elements concerning the center of mass,
$\widetilde{T}_N^{ins}$, the potential energy of its elements
inside the subsystem, $\widetilde{U}_N^{ins}$, and finally it is
an energy of interaction with other subsystems. We call,
$E_N^{ins}=\widetilde{T}_N^{ins}+\widetilde{U}_N^{ins}$, the
binding energy. In absence of external forces, the energies
$T_N^{tr}$ and $E_N^{ins}$ are motion integrals.

The kinetic energy, $T_n^{tr}$, is a functions of velocity of the
subsystems. The binding energy is determined by relative
velocities and distances between elements. Therefore we will make
such replacement of variables in which the
binding energy and kinetic energy of motion of the system will be
written down through independent variables in the equations of
motion. Such variables are relative velocities and velocity of
motion of the center of mass.

Let us assume for simplification that the system is prepared in a
nonequilibrium state of in such way that it can be divided into
two subsystems, both of them in equilibrium. The equations for the
energy exchange for two interacting subsystems have the forms
[Somsikov, 2004b, 2004c]:

\begin{equation}
{LV_L\dot{V}_L+{\sum\limits_{j=i+1}^L}\sum\limits_{i=1}^{L-1}\{v_{ij}
[\frac{\dot{v}_{ij}}{L}+\frac{\partial{U}}{\partial{r_{ij}}}]\}=
-\sum\limits_{{j_K}=1}^K}\sum\limits_{{i_L}=1}^{L}v_{i_L}
\frac{\partial{U}}{\partial{r_{{i}_{L}{j}_{K}}}} \label{eqn8}
\end{equation}

\begin{equation}
{KV_K\dot{V}_K+{\sum\limits_{j=i+1}^K}\sum\limits_{i=L+1}^{K-1}\{v_{ij}
[\frac{\dot{v}_{ij}}{K}+\frac{\partial{U}}{\partial{r_{ij}}}]\}=
-\sum\limits_{{j_K}=1}^K}\sum\limits_{{i_L}=1}^{L}v_{j_K}
\frac{\partial{U}}{\partial{r_{{i}_{L}{j}_{K}}}} \label{eqn9}
\end{equation}

Here we take ${LV_L+KV_K=0}$, ${V_L}$ and ${V_K}$ are the
velocities of the center of mass for the subsystems; $L$ and $K$
are the number of elements in the subsystems; ${v_{ij}}$ are the
relative velocities between the $i$ and $j$ elements; ${L+K=N}$.
Masses of elements are accepted to $1$. The sub-indexes denote to
which subsystems some elements belong.

The first term in the left side Eqs. (8, 9) respectively expresses
the rate of change of kinetic energy for the subsystems,
${T^{tr}}$. The second term is related to transformation of
binding energy for the subsystems, ${E^{ins}}$.

The right side in the Eqs. (8, 9) determine the energy of
subsystems interaction. The interaction is a cause of the kinetic
energy transformation of the subsystem motion, ${T^{tr}}$, into
the binding energy.

In case ${L=K}$ the equation of motion for one of the subsystem
can be deduced from Eqs. (8, 9). It takes the form [Somsikov,
2004c]:
\begin{equation}
{\dot{V}_L=- {\frac{1}{{V_L}}L}
{\sum\limits_{j=i+1}^L}\sum\limits_{i=1}^{L-1}\{v_{ij}
[\frac{\dot{v}_{ij}}{L}+\frac{\partial{U}}{\partial{r_{ij}}}]\}
-\sum\limits_{{j_K}=1}^K}\sum\limits_{{i_L}=1}^{L}
\frac{\partial{U}}{\partial{r_{{i}_{L}{j}_{K}}}} \label{eqn10}
\end{equation}

Eq. (10) determines the generalized force, which in its turn
depends not only on the coordinates but on the velocities as well.
So, the dependence of the generalized forces on the velocities of
the elements is now explicitly shown. The dependence of the
generalized forces on velocities removes the restriction on
irreversibility, superimposed by the Poincare's theorem about
reversibility [Zaslavsky, 1999] because this theorem is convenient
only to the potential forces.

In agreement with Eq. (10), we can suggest the next explanation of
the equilibration. The kinetic energy of relative motion of
interacting subsystems can be transformed in the potential energy
of their interaction and in their binding energy due to the work
of the generalized forces. As a result the relative velocities of
the subsystems are decreasing. Transformation into the potential
energy is reversible because potential energy depends only on
coordinates. But the binding energy depends both on coordinates,
and elements velocities. Its increasing can occur due to
increasing of the kinetic energy of the subsystems elements. The
binding energy cannot come back into the kinetic energy of the
subsystem motion because preservation by a subsystem of a full
momentum. Therefore the equilibration is caused by an opportunity
of transformation of the subsystem motion energy into the binding
energy and impossibility of the reversible transformation.

If the system is in the equilibrium state than the relative
velocities of subsystems and the energy flow between them are
equal to zero for any splitting [Somsikov, 2004c; Landau {\&}
Lifshits, 1976]. The equilibrium state is stable. From the
physical point of view the stability of an equilibrium state for
mixed systems is caused by aspiration to zero of the generalized
forces arising at a deviation of system from equilibrium
[Somsikov, 2004a; Rumer {\&} Ryvkin, 1977]. Hence, the system,
having come to equilibrium, never leaves this state.

Earlier we have shown [Somsikov, 2004c] that mixing provides
aspiration of the generalized force to zero. This conclusion  has
been proved in mathematical way. The mixing property has been used
for transition from summation to integration on impact parameters
of colliding disks. How it follows from the Eq. (10), this role of
the mixing in systems of potentially interacting elements is
similar.

The eq. (10) is transformed into the Newton equation in three
cases: in equilibrium state; if subsystems can be taken as a hard
without internal degrees of freedom; if the subsystem is
conservative.

\date{\it{4. Thermodynamics and classical mechanics}} \maketitle

Let us now consider the essence link between thermodynamics and
classical mechanics. It is easy to see the analogy between the
Eqs. (8-10) and the basic equation of thermodynamics [Rumer {\&}
Ryvkin, 1977]:

\begin{equation}
{dE=dQ-{PdY}} \label{eqn11}
\end{equation}

Here, according to common terminology, $E$ is internal energy of a
subsystem; $Q$ is thermal energy; $P$ is pressure; $Y$ is volume.

The energy change of the selected subsystem is due to the work
made by external forces. Therefore, the change in full energy of a
subsystem corresponds to $dE$.

The change of kinetic energy of motion of a subsystem as the
whole, $dT^{tr}$, corresponds to the term ${PdY}$. Really,
${dT^{tr}=VdV=V\dot{V}dt=\dot{V}dr=PdY}$

Let us determine, what term in Eq. (11) corresponds to the change
of the binding energy in a subsystem. As follows from virial
theorem [Landau {\&} Lifshits, 1973], if the potential energy is a
homogeneous function of second order of the radiuses-vectors, then
${\bar{E}^{ins}=2\bar{\tilde{T}}^{ins}=2\bar{\tilde{U}}^{ins}}$.
The line denotes the time average. Earlier we obtained that the
binding energy, ${E^{ins}}$, increases due to contribution of
energy, ${T^{tr}}$. But the opposite process is impossible.
Therefore the change of the term $Q$ in the Eq. (11)  corresponds
to the change of the binding energy ${E^{ins}}$.

Let us consider the system near to equilibrium. If the subsystem
consist of ${N_m}$ elements, the average energy of each element
becomes, ${\bar{E}^{ins}={E}^{ins}/N_m=\kappa{T}_0^{ins}}$. Now
let the binding energy increases with ${dQ}$. According to the
virial theorem, keeping the terms of the first order, we have:

${dQ\approx{T}_0^{ins}[d{E}^{ins}/{T}_0^{ins}]
={T}_0^{ins}[{dv}/{v_0}]}$, where ${v_0}$ is the average velocity
of an element, and ${dv}$ is its change. For subsystems in
equilibrium, we have ${dv/v_0\sim{{d\Gamma_m}/{\Gamma_m}}}$, where
${\Gamma_m}$ is the phase volume of a subsystem, ${d\Gamma_m}$
will increase due to increasing of the subsystem energy on the
value, ${dQ}$. By keeping the terms of the first order we get:
${dQ\approx{T}_0^{ins}d\Gamma_m/\Gamma_m={{T}_0^{ins}}d\ln{\Gamma_m}}$.
By definition ${d\ln{\Gamma_m}=dS^{ins}}$, where ${S^{ins}}$ is a
subsystem entropy [Landau {\&} Lifshits, 1976; Rumer {\&} Ryvkin,
1977]. So, near equilibrium we have
${dQ\approx{T}_0^{ins}dS^{ins}}$.

\date{\it{5. Relation between entropy and generalized forces}} \maketitle

Let us consider the relation of the generalized field of forces
with entropy. According to the results obtained here, and also in
agreement with [Landau {\&} Lifshits, 1976] the equilibrium state
of the system is characterized by absence of energy of relative
motion, ${T_m^{tr}}$, of subsystems. I.e. energy ${T^{tr}}$, as a
result of the work done by the generalized forces, will be
redistributed between the subsystems into the binding energy. This
causes an increase of entropy. When the relative velocities of the
subsystems go to zero, the system goes to equilibrium. So, the
entropy increasing, $\Delta{S}$, can be determined as follows :

 \begin{equation}
{{\Delta{S}}={\sum\limits_{l=1}^R{\{{\frac{m_l}{E^{m_l}}}
\sum\limits_{k=1}^{m_l}\int{\sum\limits_s{{F_{ks}}^{m_l}}{dr_k}}\}}}}\label{eqn12}
\end{equation}

Here ${E^{m_l}}$ is the kinetic energy of subsystem; ${m_l}$ is
the number elements in subsystem ${"l"}$; ${R}$ is the number of
subsystems; ${s}$ is number of the external disks which collided
with internal disk ${k}$. The integral is determining the work of
the force ${F_{ks}^{m_l}}$ during the relaxation to equilibrium.
It is corresponds to phenomenological formula Clauses for entropy
[Landau {\&} Lifshits, 1976]. So, Eq. (12) deduced entropy from
the first principles of the classical mechanics through the
generalized force. Therefore we can use this equation for
analyzing different kind of entropy [Klimontovich, 1995; Tsallis,
Baldovin, Cerbino, Pierobon, 2003]. Inherently Eq. (12)
corresponds to the formula for entropy:
${S=\sum\limits_{l=1}^R{S_l(E_l^{ins}+T_l^{tr})}}$, (see, [Landau
{\&} Lifshits, 1976]). In fact, if ${E_l^{ins}\gg{T_l^{tr}}}$,
then
${dS=\sum\limits_{l=1}^R\frac{\partial{S_l}}{\partial{T_l^{tr}}}{dT_l^{tr}}}$
which corresponds to Eq.(12). Thus, the Eq. (12) connects dynamic
parameter, which is force acting on a subsystem, with entropy,
which is a thermodynamic parameter. I.e. this formula established
the connection between parameters of classical mechanics and
thermodynamic parameters.

\date{\it{6. Conclusion}} \maketitle

Process of the solution of a problem of irreversibility has led to
the concepts of the generalized Liouville equation, the
generalized force, and also to the equation of motion for
subsystems. The Liouville equation and the motion equations of
subsystems allow to find a necessary condition of occurrence of
irreversibility and to show that the forces of interaction of
subsystems in nonequilibrium systems are non-potential. From the
equation of motion of subsystems follows, that the work of the
generalized forces carries out irreversible transformation of
energy of the relative motion of subsystems in their binding energy.
It determines relation of classical mechanics with thermodynamics.
Thus, the solving of a problem of
irreversibility leads to the solving of a problem of a
substantiation of thermodynamics.

The essence of the mechanism of irreversibility offered by us
consists in the following. The energy of relative motion of
subsystems of nonequilibrium system is distinct from zero. As
against kinetic and potential energy of element, it depends on
character of function of distribution of these elements. As it
follows from the equation of motion of subsystems, this energy as
a result of work of the generalized forces is capable to be
transformed by irreversible way to the binding energy of
subsystems. Therefore the reduction of energy of motion of
subsystems leads to reduction of the generalized forces.

Transformation of energy of motion of subsystems to their binding
energy occurs as a result of increase in energy of elements of a
subsystem. Irreversibility of such transformation of energy is
provided with impossibility of increase in velocity of motion of a
subsystem due to the binding energy. This impossibility is
connected with the spherical symmetry of function of distribution
of velocities of elements of a subsystem and preservation of this
symmetry. The process of transformation of energy of relative
motion of subsystems goes until this energy completely will pass
to the binding energy. I.e. the energy of the motion of subsystems
determined by function of distribution of velocities of system
goes on entropy increase of systems. Therefore this energy
determines the rate of the entropy deviation from its equilibrium
state. It corresponds to the fact that the total work of the generalized
forces on all subsystems is equal to zero. Thus, in a basis of the
offered mechanism of irreversibility the state of impossibility of
infringement of symmetry of function of distribution of velocities
of a subsystem due to internal forces is used.

We can find the interrelation of classical mechanics with thermodynamics
with the help of the motion equation of the subsystems.
This interelation is caused by that the kinetic energy of relative motion of
subsystems can be transformed not only to the potential energy of
their interaction, but also into their binding energy. Therefore
the work on the closed contour is not equal to zero. It
corresponds to the first law of thermodynamics.

The generality of the offered mechanism of irreversibility is
determined by a generality of the chosen model of nonequilibrium
system. As in this mechanism the interaction of elements of systems
determine the process of an establishment of equilibrium. Therefore
this mechanism is not applicable for systems similar to ideal gas and Brownian
particles in which the establishment of equilibrium is caused by
interaction with an environment and therefore determined by the
probability laws [Smoluchowski,1967].

Within the framework of the
problem put here, the chosen model of nonequilibrium system and
methods of its analysis do not contradict to the classical
mechanics. Similar model have a place in the real world. It gives
the basis to consider, that the connection found here between
classical mechanics and thermodynamics qualitatively corresponds
to the validity.

\end{document}